# From Ideation to Implications: Directions for the Internet of Things in the Home

**Albrecht Kurze**
Chair Media Informatics
Chemnitz University of Technology, Germany
Albrecht.Kurze@informatik.tu-chemnitz.de

**Arne Berger**
Chair Media Informatics
Chemnitz University of Technology, Germany
Arne.Berger@informatik.tu-chemnitz.de

**Teresa Denefleh**
Chair Media Informatics
Chemnitz University of Technology, Germany
Teresa.Denefleh@informatik.tu-chemnitz.de

## ABSTRACT

In this paper we give a brief overview of our approaches and ongoing work for future directions of the Internet of Things (IoT) with a focus on the IoT in the home. We highlight some of our activities including tools and methods for an ideation-driven approach as well as for an implications-driven approach. We point to some findings of workshops and empirical field-studies. We show examples for new classes of idiosyncratic IoT devices, how implications emerge by (mis)using sensor data and how users interacted with IoT systems in shared spaces.

## INTRODUCTION

The Internet of Things has arrived in the home. More and more smart devices in the home are sensor-equipped and connected. These connected devices offer (or at least promise) new chances and new possibilities, e.g. more comfort, more security, more safety, and more efficiency. Nevertheless, these applications are not necessarily (all) what the users want and what the IoT is limited to. We believe that the IoT will not only become an automated IoT but also a social IoT. Everyday objects will be used for subtle communication, as emotional devices or even as companions in everyday life. These objects will probably be more than ordinary smart screens and speakers.

In this position paper we give a brief overview of our approaches and ongoing work for future directions of the IoT with a focus on the IoT in the home. We believe it takes two approaches for new directions in the IoT and both approaches benefit of participatory HCI research:

1) Ideation-driven approach: We believe better products and services need better ideation and design processes. Designing for the IoT design space is a challenge for design, because the IoT combines the tangible material of things and the intangible material of services and networks. To support design activities within this space, a plethora of IoT design methods exists. We introduce in the second section our tools and methods for ideation for the IoT, how we used them in workshops and present some of our workshop findings that show how future IoT devices might be smart but also idiosyncratic.

2) Implications-driven approach: The same time the IoT introduces new chances and possibilities it might also introduce some new risks and threats that the users are not aware of. Therefore, it needs a view on the implications of the IoT, especially on ethical and privacy issues in relation to the user data. Often it is not enough to stay on a conceptual level. Instead it needs the users to experience the possibilities of the technology to reflect critically on implications. We introduce in the third section our tools and methods that we used in empirical field studies to research how users deal with the chances but also risks of the IoT.

## IDEATION-DRIVEN APPROACH: FROM POETIC TOY PETS TO AUTOMATED SYSTEMS

We developed tools and methods for ideation for the IoT. *Cards'n'Dice* is a combination of the *Loaded Dice* [1,9,10] and sets of cards. The *Loaded Dice* (figure 1) are a tool for designing scenarios revolving around the IoT. They consist of two wirelessly connected dice; a sensor die that can measure six different qualities and an actuator die that has six different outputs. The top face is the active face so that just the turn of a die results in the creation of a new input-output combination. A number of methods can be created with the *Loaded Dice* using different card sets. We created a card set for scenario development (described in [2]) and a new one for focusing on micro-interactions with sensor and actuator enabled mundane objects. Our tools and methods offer a playful way for lay people to explore the basic principles of the IoT and design an IoT device in workshops without requiring expert technical knowledge on the IoT.

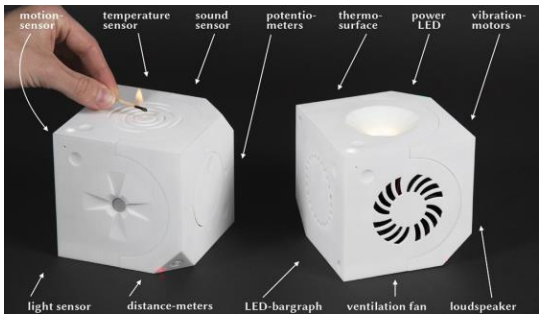

Figure 1: *Loaded Dice* for IoT ideation. Left: sensor die, right: actuator die.

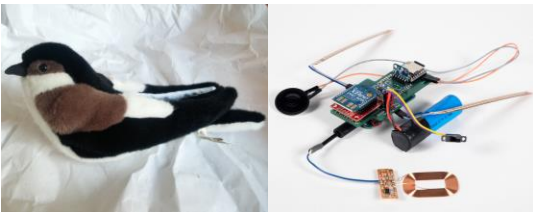

Figure 2: *WhetherBird.* Tells whether there was rain or not. Outside a toy pet (northern house martin), inside a smart connected IoT device

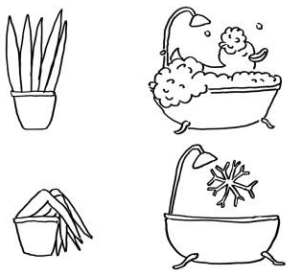

Figure 3: The *automated rent debtor* principle that participants developed. Overdue rent debts or overdue housework tasks have consequences: no more hot water for debtors.

We used our tools and methods in a number of workshops, e.g. with users that have special needs (blind and visually impaired) [9], with elderly people in our living lab [4], and in special constellations of living in a community housing [2].

In nearly all cases the participants in our workshops started to ideate IoT devices and systems that go far beyond the paradigms of smart screens and speakers. Inspired by the possibilities of what a smart combination of sensors and actuators in connected devices can do the participants often created systems that demand a variety of innovative and embedded actuators for the IoT.

An ever-repeating finding is the idiosyncratic ideation. The users tend to design devices and systems that are emotional, sometimes even poetic and that embed poetic interactions in mundane objects. Often these are idiosyncratic qualities with the potential of different meanings for different people or no meaning for outsiders at all. Participants brought up the idea of the *WhetherBird* (figure 2), a system that uses simple IoT sensor data loaded with individual meaning in a poetic IoT device [9]. The *WhetherBird* provides a weather 'aftercast' for blind people to let them choose the right shoes when leaving the house in the morning. Depending on whether there *was* rain in the night it modulates the singing of a small toy pet in shape of a bird. Dry weather and ground make the bird sing friendly, bad weather and wet ground make it sing sadly. Other examples in the same direction of new interaction schemes and idiosyncratic use of sensors and actuators are described in [2].

Our participants also ideated automated IoT scenarios. One scenario, the *automated rent debtor* (figure 3), directly addresses financial status data as input for decisions and actions in a community housing [2]. A common issue there is, that flat mates do not pay their rent on time. The participants ideated a gradually escalating penalty system. Depending on the overdue time and amount of money the debtors first receive a simple text messages on a phone. Later the automated IoT system escalates the actions to emphasize the admonition, e.g. the debtor would not be able to take hot showers. While such a system would not be a problem from a technical point of view, it does not mean that it would be free of problems or implications. During the ideation process of the automated rent debtor the participants started to reflect on the implications that such a system would create and that it would be inhumane.

This brings us to the implications-driven approach for new IoT directions.

## IMPLICATIONS-DRIVEN APPROACH: FROM SIMPLE IOT SENSOR DATA TO WICKED IMPLICATIONS

A purely theoretical approach to reflect on implications of the IoT is often not enough. In addition, empirical studies of the actual use of IoT devices and services are required to reflect on implications and to identify them in the first place. Therefore, we developed own tools and methods for this purpose and applied them in field studies.

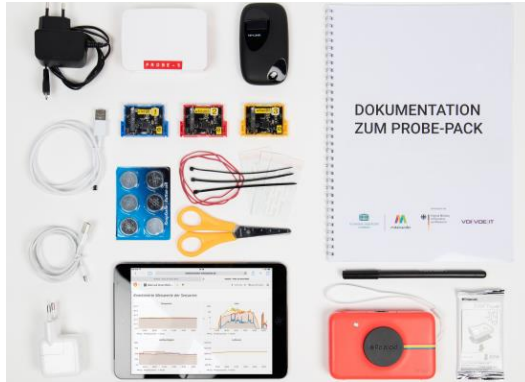

**Figure 4: Components of the *Sensing Home* kit (SensorTags, Raspberry Pi, tablet pc, etc.) .**

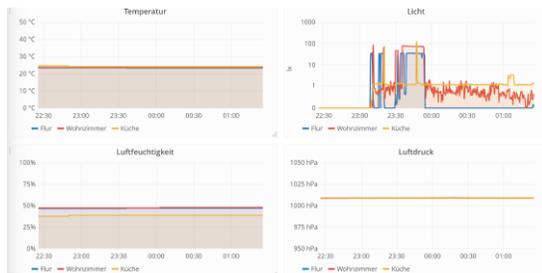

**Figure 5: Screenshot of data visualization in graphs as seen by the participants on the tablet computer (example).**

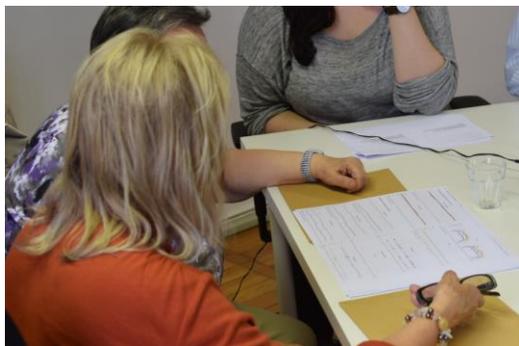

**Figure 6: *Guess the Data*, participants discussing.**

### *Sensing Home* kit and *Guess the Data* collective data work method

Smart Home IoT technologies increasingly collect seemingly inconspicuous data from homes. It might not even need a camera or a microphone to break the intimacy of a home. Even simple and at first sight mostly harmless sensors, e.g. for light, temperature or humidity, may reveal a lot about the people, their presence or absence, the daily routines, maybe even about their behavior and their preferences [7,11,13]. Understanding both sides, chances and risks, requires awareness and agency in the usage of these devices. However, little is known how users account collectively and collaboratively for simple sensor data from their homes and those of others.

For our participatory research on this topic, we developed the *Sensing Home* kit [2,3]. It includes all components (figure 4) to quickly setup the collection of simple IoT sensor data and to let participants view the collected data and reflect on it. We decided for simple sensors and chose the TI SensorTag as a suitable base. It offers battery operation, wireless communication and a large number of different simple sensors (lux-, thermo-, hygro-, baro-, accelero-, gyro- and magnetometer) at a low price (USD 30). The SensorTags are small and lightweight enough to attach them nearly everywhere. A Raspberry Pi 3 works as an edge gateway. It connects the wireless sensors and forwards the data to a server in our department for storage and visualization. A preconfigured tablet computer gives participants full access to their live and historic sensor data (figure 5).

We conceptualized and field-tested our two-fold method *Guess the Data*. Our method adapts and extends the concept of data work [6,7]. First, we installed the sensors in the homes for two weeks. During that time the participants collected and explored their sensors data in an individual data work enabled by our system. Second, we invited the participants of different households to group discussions. In a collective data work we presented them their anonymized sensor data in form of printed graphs (figure 6). We wanted the participants to guess the data: what they think, from which household the data is, and what they recognize in the data. The collective data work also acted as a catalyst for further discussion and reflection on the data and its implications.

Findings of our study, conducted in nine households, show how users make sense and use of simple sensor data. Besides the already known practices of interpretation of sensor data [6,7,13] we identified a collective dimension of situating IoT sensor data from the home in social contexts.

As our method gives the participants the ability to use the sensor data during the data collection stage for own purposes we found struggling ethical and privacy issues in this use. We could reveal the potential of misuse of sensor data as means of power and for surveillance against other family members [12].

In the study we realized a concept that we first presented at a previous CHI workshop [8]. Nevertheless, at this time we were not aware of the sometimes wicked implications that came up. One of the wicked implications is that the less abstract and more concrete threat for an individual's privacy might not come from an anonymous third party Big Brother but a big mother or other member of the household. We see our method as a powerful participatory HCI method for the IoT in the home to reveal these wicked implications.

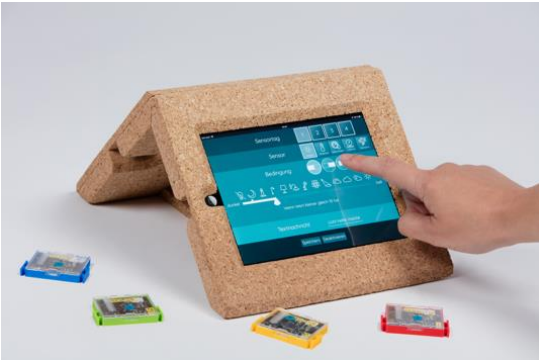

**Figure 7:** *Sensorstation* **input screen side and color-coded SensorTags.**

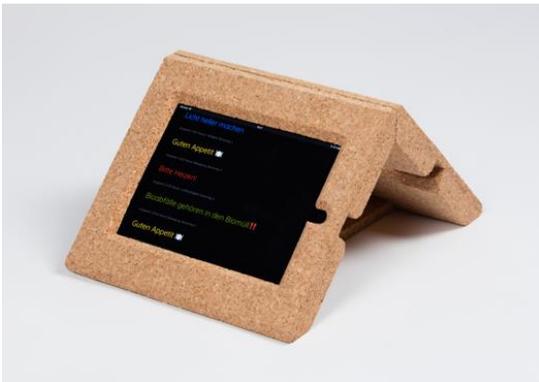

**Figure 8:** *Sensorstation* **output screen side with message stream. The notifications are time-stamped and color-coded according to the assigned SensorTag.**

### *Sensorstation*: Towards the implications of use of shared data in shared spaces

By using the *Sensing Home* kit, we learnt how to design critical smart devices that teach about the power of seemingly harmless sensors and allow to investigate awareness and especially agency in using IoT sensor devices and IoT applications. Based on the *Sensing Home* kit we developed *Sensorstation* (figures 7 and 8), a research product that empowers participants to explore the use of simple IoT sensor data [5]. It uses the same simple IoT sensors in combination with a stationary device (the station) that consists of an input screen for configuration and an output screen for displaying notifications. The concept of *Sensorstation* introduces an abstraction layer to the raw data. In a configuration interface, users can set conditions, e.g. events, thresholds or range, for the individual color-coded sensors and sensor functions (motion, light, temperature, humidity, air pressure) and assign a short message (text and/or Emoji) to this condition. Every time a condition is met a notification is sent to a shared message stream that is displayed on the output screen. The stationary part is intended to be placed in a central position in the flat, accessible and readable for all inhabitants, e.g. on the kitchen table, while the sensors can be used everywhere in the flat.

With a first deployment we were interested in exploring what sensor applications inhabitants would develop and what effects this can have on people living together. We deployed *Sensorstation* in a shared apartment for 19 days to investigate the effects of smart sensors and data in the context of shared and private spaces. The stationary part was deployed in the kitchen, a shared space in the flat, for shared usage (configuration as well as message stream display). The SensorTags were assigned (1:1) to the flat-mates for personal use. After the deployment stage we performed a group discussion in the flat.

Our findings show that the participants explored a number of different smart home applications with *Sensorstation*. These ranged from sensor applications for creating positive connections within the group, over self-monitoring, to control over others as well as reward systems and penalties. *Sensorstation* in combination with the group discussion not only provided insight into the use of sensor data in the shared apartment, but also into the underlying usage motives. The influence of the technology on the group and possible chances and conflict potentials in the use of sensor technology lies in the implications of shared sensor data in shared spaces. We subsequently argue that design has an obligation to consider smart technology that acknowledges boundaries and that provides a negotiation space to configure agency within these.

### SUMMARY AND CONCLUSION

We discussed two possible approaches for future IoT directions and illustrated them along our work portfolio. We have shown the potential of an ideation-driven approach, which we pursue with our own tools and methods and presented some examples of ideated idiosyncratic IoT devices and automated IoT systems for the home. Other great potential for future IoT directions lies in the examination of the implications that IoT devices and systems create, especially in a potentially sensitive context like 'the home'. We presented our tools and methods that we use in empirical field studies and pointed to some implications, especially for ethical and privacy issues.

Kurze et al.    6

## ABOUT US

The contributors work in an interdisciplinary team called *Miteinander* (German for "together") at Chemnitz University of Technology, bringing together competence in design, computer science, social sciences and engineering. We investigate participation and co-creation for smart connected IoT technology in the home within the realm of demographic change and community work.

**Albrecht Kurze:** Albrecht is a computer scientist. His research interests are networking aspects in all flavors. In his interdisciplinary PhD thesis he quantified the relationship between Quality of Service and Quality of Experience for mobile services with about 300 participants. At *Miteinander* he is in charge of the engineering for the IoT. The interdisciplinary approach of the team shifted his focus and view - once again - away from only looking on the technology - much more to the users and the implications that we create with the smart connected technology.

**Arne Berger:** Officially a computer scientist with a doctorate in engineering, Arne's research takes an inter- and transdisciplinary research through design approach at the intersection of design ethnography and interaction design. Arne is the principal investigator of the Interaction Design Research Lab team *Miteinander*. His research focuses on early stages of design processes. He is mainly interested in how meaningful participation can be initiated, how people can be empowered and engaged to collaboratively create possible futures together. These days he is also turning towards more ethnographical design research work.

**Teresa Denefleh**: Teresa is a computer scientist. She is concerned with fields of application and social effects of sensor technology in living areas. Her research interest lies in the interaction between humans with the help of sensors and IoT artefacts. In her interdisciplinary master thesis she developed and investigated a research tool for sensor exploration in the context of a residential community (*Sensorstation*). After her master thesis she joined *Miteinander* in autumn 2018.

## ACKNOWLEDGMENTS

This research is funded by the German Ministry of Education and Research (BMBF) under grant number FKZ 16SV7116. (c) Photographs by team *Miteinander*.